\newcommand{\e}{\mathrm{e}}
\newcommand{\Deltapw}{\Delta_{\mathrm{pw}}}
\newcommand{\psipw}{\psi_{\mathrm{pw}}}
\newcommand{\x}{\vec{x}}

\newcommand{\q}{\vec{q}}

\newcommand{\qp}{\vec{q}\,'}

\newcommand{\bsub}{\begin{subequations}}
\newcommand{\esub}{\end{subequations}}
\newcommand{\beq}{\begin{equation}}
\newcommand{\eeq}{\end{equation}}
\newcommand{\beqa}{\begin{eqnarray}}
\newcommand{\eeqa}{\end{eqnarray}}
\newcommand{\beql}{\begin{subequations}\begin{eqnarray}}
\newcommand{\eeql}{\end{eqnarray}\end{subequations}}
\documentclass[aps,prl,twocolumn,groupedaddress,amsmath,showpacs,nofootinbib]{revtex4}
\usepackage{amsmath} 
\usepackage[pdftex]{graphicx}
\usepackage{float}   % per posizionare meglio le figure con [H]
\usepackage{verbatim}  %multi-line comment \begin{comment}....\end{comment}
\usepackage[usenames]{color}	% comments in colors
\definecolor{darkgreen}{rgb}{0,0.4,0} 
\begin{document}
\title{X-entanglement of PDC photon pairs}
\author{E.~Brambilla, L.~Caspani, O. Jedrkiewicz, L.~A.~Lugiato and A.~Gatti}
\email[Corresponding author: A. Gatti:]{alessandra.gatti@mi.infn.it}
\affiliation{INFM-CNR-CNISM, Dipartimento di Fisica e Matematica,
Universit\`a dell'Insubria, Via Valleggio 11, 22100 Como, Italy}
\begin{abstract}
We investigate the spatio-temporal structure of the bi-photon entanglement in 
parametric down-conversion (PDC) and we demonstrate its non-factorable X-shaped geometry. 
Such a structure gives access to the ultra-broad bandwidth of PDC, and can be exploited 
to achieve a bi-photon temporal localization
in the femtosecond range. This extreme localization is 
connected to our ability to resolve the photon positions in the source near-field. The non factorability opens the possibility of tailoring the temporal entanglement by acting on the 
spatial degrees of freedom of twin photons.
\end{abstract}
\maketitle
Parametric down-conversion (PDC) is probably the most efficient and widely used source of entangled photon pairs, which has been employed
in several successful implementations of quantum communication and information schemes.
At the very heart of such technologies lies the quantum interference
between photonic wave functions, which depends crucially on the spatio-temporal mode structure of the photons.
In this work, the issue of controlling and tailoring the bi-photon spatio-temporal structure is addressed from a  peculiar and novel point of view, that is, the non factorability in space and time of the PDC bi-photon entanglement. The idea comes from the context of nonlinear optics, where recent studies \cite{conti2007} outlined how in nonlinear media the angular dispersion relations impose a hyperbolic geometry involving
both temporal and spatial degrees of freedom in a non-factorable way. The wave object that captures such a geometry
is the so-called X-wave (the X being formed by the asymptotes of the hyperbola), which is a localized and  propagation-invariant wave-packet,  non separable in space and time. The statistical counterpart of the X-wave was recently showed to emerge in the X-shaped structure of the classical first order coherence function \cite{jedr2007}. 
\par
In this work, we turn our attention to the genuine quantum properties of PDC, 
and we adopt the X-wave picture for investigating the spatio-temporal structure of the two-photon entangled state. 
With few exceptions \cite{atature2002,gatti2004}, investigations on the quantum state of PDC have been performed to date 
mostly either in a purely temporal \cite{HOM,law2000,grice2001} 
or spatial \cite{rubin1996,law2004,gatti2008} framework. Our approach, based on the non-factorability in space and time of the state, will point out relevant elements of novelty, as the possibility of tailoring the temporal bandwith of the bi-photons by manipulating their spatial degrees of freedom.  In particular, by resolving their near-field positions, we will show that the X-structure opens the access to an ultra-broad bandwidth entangled photonic source, with a temporal localization in the femtosecond range. Such an extreme localization can be applied to increase the sensitivity of high precision measurements in the time domain (e.g. in  the protocol of clock syncronization \cite{giovannetti2001} or of quantum optical coherence tomography \cite{abouraddy2002}).
Our results compare with recent findings reported in \cite{nasr2008}, where a $\sim 7fs$ Hong-Ou-Mandel dip was observed
through the use of a quasi-phase-matched nonlinear grating. 
\par
We shall focus on type I PDC, in the low-gain regime where single pairs of photons can be detected by coincidence counts. We remark that the X-structure of entanglement is a general feature of PDC, present also in  type II and in the high-gain regime \cite{caspani2008}.
\par
The model is basically the same as in \cite{gatti2004, gatti2008}. A quasi-monochromatic and coherent pump field propagates along the direction $z$ inside a slab of nonlinear $\chi^{(2)}$ crystal of length $l_c$. $\hat A_p(\x,t,z)$, $ \hat A_s (\x,t,z)$ denote the envelope operators of the pump and the down converted signal field, of central frequencies $\omega_p$ and $\omega_s=\omega_p/2$, respectively. Here $\x=(x,y)$ labels the  transverse coordinates, while $t$ is time. We next pass to the Fourier domain: $\hat A_i (\q,\omega, z) = \int \frac{d^2 \q}{2\pi} \int  \frac{d\omega}{\sqrt{2\pi}}
 \hat A_i(\x,t,z) \e^{-i \q\cdot \x + i \omega t}$, 
$i=s,p$ and extract the fast variation due to their linear propagation along the crystal slab:
\beq
\hat A_i(\q,\omega, z)= \e^{i k_{iz} (\q,\omega) z} \hat a_i( \q,\omega,z) \quad i=s,p \, ,
\label{ai}
\eeq
where $
k_{iz} (\q,\omega) = \sqrt{ k_i^2(\q,\omega) -q^2}
$
is the $z$-component of the wave vector of the i-th field, $k_i(\q,\omega)$ being the wave number at frequency $\omega$, which for the extraordinary wave depends also on the propagation direction, identified by the transverse wave-vector $\q$. The fields $\hat a_i$ defined in this way have a slow variation along the crystal, arising only from the nonlinear interaction. In the low-gain regime we can assume that the pump is undepleted by the nonlinear interaction, so that $\frac{d}{dz} \hat a_p(\q,\omega,z)=0$,  and  substitute its field operator by a c-number field $ \alpha_p(\q,\omega)$. In this way, the pump evolution along the crystal is described by  $\mathcal{A}_p (\q,\omega,z) = \e^{i k_{pz} (\q,\omega) z} \alpha_p( \q,\omega)$.  For the down-converted signal field, its propagation along the crystal  is described by the equation \cite{gatti2004}:
\beqa
\frac{\partial \hat a_s(\q,\omega,z)}{\partial z} &=&
    \frac{g}{l_c} 
    \int \frac{d^2 \qp}{2 \pi} \int \frac{d\omega'}{\sqrt{2\pi}}
         \left[ \alpha_p(\q+\qp,\omega+\omega') \right.  \nonumber \\
&\times& \left.\hat a_s^{\dag}(\qp,\omega',z)
         e^{-i\Delta(\q,\omega,\qp,\omega')z} \right] \;,
\label{waveq} 
\eeqa
where $g$ is the dimensionless parametric gain, proportional to the second-order $\chi^{(2)}$ susceptibility, to the crystal length $l_c$ and to the pump peak value (the pump field has been normalized to its peak value). 
The phase-mismatch function 
\beq
\Delta(\q,\omega,\qp,\omega')=k_{sz}(\q,\omega)+k_{sz}(\qp,\omega')-k_{pz}(\q+\qp,\omega+\omega')  
\label{mismatch}
\eeq
 determines how efficiently a pump photon with transverse wave-vector $\q+\qp$ and frequency $\omega_p+\omega+\omega'$ is down-converted into a pair of photons with transverse wave-vectors $\q,\qp$ and frequencies  $\omega_s+\omega$, $\omega_s+\omega'$: the smaller its modulus, the higher the probability that such an elementary process occurs. 

In most experiments in the low gain regime, the quantity of primary interest is the two-photon correlation, also called {\em bi-photon amplitude}. 
We shall study this quantity in the spatio-temporal domain, in a plane at the output face of the crystal ({\em near-field region}), that is, we focus on 
\beq
\label{correl1}
\psi(\x,t,\x',t')=\langle \hat A_s(\x,t,l_c)\hat A_s(\x',t', l_c)\rangle \; .
\eeq
In the low gain limit considered in this work its square modulus $|\psi(\x,t,\x',t')|^2$ is proportional to the two photon coincidence rate
$G^{(2)}(\x,t,\x',t)$, which gives the joint probability distribution of finding two photons  in position $\x$ at time $t$ and position $\x'$ at time $t'$, respectively. 
\par
For small gains ($g\ll 1$), the propagation equation (\ref{waveq}) can be solved
perturbatively up to first order in $g$, obtaining the following expression for the bi-photon amplitude in the Fourier domain:
\beq
\begin{split}
\label{correl2}
\langle & \hat A_s(\q_1,\omega_1,l_c)\hat A_s(\q_2,\omega_2,l_c)\rangle=  g
 {\mathcal A}_p(  \q_1+\q_2,\omega_1+\omega_2,l_c) \\
& \times      (2\pi)^{-3/2} 
 e^{i\frac{\Delta  (\q_1,\omega_1,\q_2,\omega_2)   l_c}{2}}    
\text{sinc}\left[\frac{ \Delta (\q_1,\omega_1,\q_2,\omega_2)l_c}{2}\right] 
\end{split}
\eeq
In the literature the same quantity is usually derived through a perturbative evaluation of the two-photon state vector (see e.g. \cite{atature2002}).
\\
In order to simplify our results, we consider the limit of a nearly plane-wave and monochromatic pump, i.e, a pump of duration $\tau_p$ and waist $w_p$ large enough, so  that the dependence of the phase-mismatch  $\Delta$ on $ \q_1+\q_2$ and $\omega_1+\omega_2$ (the pump variables) can be neglected. 
It can be shown \cite{caspani2008} that such an approximation holds when  $w_p$ and $\tau_p$ are much larger 
than the spatial walk-off and the temporal delay due to group velocity mismatch, respectively, experienced by the signal and the pump after crossing the crystal. Typical values are $\sim 300 \, \mu$m and $\sim 2$ps, as in the example 
of a 4 mm $\beta$-barium-borate (BBO) crystal cut for degenerate type I PDC at 352 nm.  
Provided that we are in such conditions, the bi-photon amplitude (\ref{correl1}) at the crystal output face takes the factorized form:
\beq
\label{correl3}
\psi(\x,t,\x',t')=
\mathcal{A}_p\left(\frac{\x+\x'}{2},\frac{t+t'}{2}, l_c \right)
\psipw(\x-\x',t-t')
\eeq
where 
\bsub
\label{correl4}
\beqa
\label{correl4a}
\psipw(\x,t)&=& 
\int \frac{d^2 \q}{(2\pi)^2} \int \frac{d\omega}{2\pi} \e^{i\q\cdot\x-i\omega t} V(\q,\omega)   \\  
V(\q,\omega) &=&
g e^{i\frac{\Deltapw(\q,\omega)l_c}{2}}  
\text{sinc}\left[\frac{\Deltapw(\q,\omega)l_c}{2}\right] \, ,
\label{correl4b} \\
\Deltapw (\q,\omega) 
&=& k_{sz} (\q,\omega) +   k_{sz} (-\q,-\omega) - k_p (0,0)  
\label{deltapw}
\eeqa
\esub
is the PWP result for the field correlation function. The pump beam profile $\mathcal{A}_p(\x,t,l_c)$ acts thus as a slow modulation over the PWP correlation 
$\psipw(\x,t)$, as it can be expected in the nearly stationary and homogeneous conditions considered here.
\par
A first qualitative insight into the problem can be obtained by considering the usual quadratic expansion of the phase matching function, equivalent to adopting the paraxial and quadratic dispersion approximations.
In the case of e-oo phase matching, it takes the form \cite{gatti2004}:  
\beq
\Deltapw(\q,\omega) l_c \approx 
\Delta_0  +\frac{\omega^2}{\Omega_0^2}-\frac{q^2}{q_0^2} \, , 
\label{quadratic}
\eeq
where $\Delta_0=(2k_s-k_p) l_c$ is the collinear phase mismatch at degeneracy, $\Omega_0=\sqrt{1/k_s''l_c}$,
 $q_0=\sqrt{k_s/l_c}$, 
and we used the short-hand notation $k_s= k_s(0,0)$, $k_p= k_p(0,0)$, $k_s''= d^2k_s/d\omega^2  |_{0,0}$.  
If we extend the validity of such an approximation to the entire $(\q,\omega)$ domain, and we 
use the identity $ \e^{i p/2} \mathrm{sinc} (p/2) = \int_0^1 ds e ^{i sp}$, 
the bi-photon amplitude $\psipw(\x,t)$ 
can be recasted  in the integral form: 
\beq
\psipw(r,t)
= g\frac{q_0^2\Omega_0}{8\sqrt{\pi^3 i}} \int_0^1 
                \frac{ds}{s^{3/2}} e^{\frac{i}{4s}(q_0^2 r^2-\Omega_0^2t^2)}e^{i s\Delta_0}\;.     
\label{correl5}           
\eeq
where $r$ indicates the radial coordinate.
This 
expression clearly evidences the hyperbolic geometry of $\psipw(r,t)$:  the function is indeed  constant on the rotational hyperboloids where the argument 
\beq
H(r,t)\equiv q_0^2 r^2-\Omega_0^2t^2
\eeq 
assumes constant values.   
However, it can be easily shown that  $\psipw(H)$ goes as $1/\sqrt{|H|}$ for $|H|\rightarrow 0$, that is, 
when approaching  the asymptotes of the X-structure,  where  
$H(r,t)=q_0^2r^2-\Omega_0^2 t^2=0$. This singularity arises from the unphysical assumption that the approximation \eqref{quadratic} is valid everywhere.  
\par
\begin{figure}[h!]
\includegraphics[width=7.6cm,keepaspectratio=true]{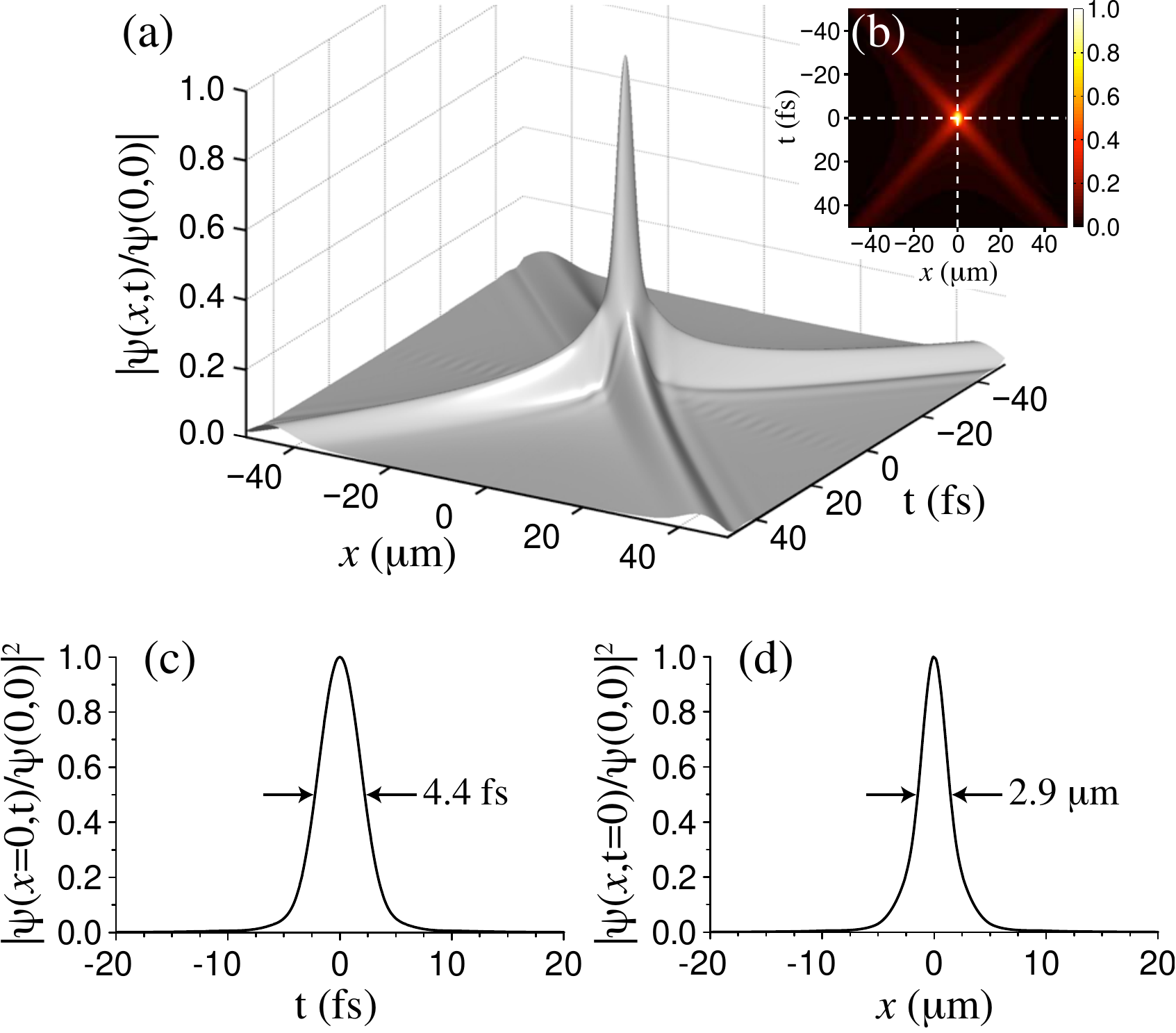}
\caption{(Color online)(a) and (b): Modulus of the bi-photon amplitude in the plane $(x,t)$, clearly displaying its X-shaped geometry (the whole 3-D plot has a rotational symmetry in space). 
(c) and (d): Cuts of the coincidence rate $ \left| \psipw(\x,t)\right|^2$ along the temporal (c) and the transverse coordinate axis (d) [indicated by the two dashed lines in frame (b)]. The width of the peaks shows the relative temporal and spatial localization of bi-photons. BBO crystal, cut at $33.436^{\circ}$ for type I PDC; $g=10^{-3}$, $\lambda_p=352$nm, and $l_c=4$mm.}
\label{fig_1}
\end{figure}

In order to obtain quantitative results we need therefore to drop the approximation \eqref{quadratic}, and to go beyond the paraxial and quadratic dispersion approximations in the evaluation of $\Delta(\q,\omega)$.  $\psipw(\x,t)$, defined by Eq.(\ref{correl4}), is hence numerically calculated 
by using the complete Sellmeier relations \cite{boeuf2000} for the refractive indexes. An 
example of our results is shown by Fig.\ref{fig_1} for the case of a type I 
BBO crystal.  Since the signal is an ordinary wave, the spatial radial symmetry of the problem can be exploited to compute  Eq.(\ref{correl4}) by means of a Fourier-Hankel transform. 
Correspondingly, the bi-photon amplitude $\psipw$ depends only on the radial coordinate $r=|\x|$. 
%%%%%%%%%%%%%%%%%%%%%%%%%%%%%% FIGURA 6 panels %%%%%%%%%%%%%%%%%%%%%%%%%%%%%%%%%%%%%%
%%%%%%%%%%%%%%%%%%%%%%%%%%%%%%%%%%%%%%%%%%%%%%%%%%%%%%%%%%%%%%%%%%%%%%%%%%%%%%%%%%%%%%%%%%%%%%%%%%%%%%%%%%%%%%%%%%%%%%%%%%%%%%%%%
A cut of this quantity in the $(x,t)$ plane  is displayed by frames (a) and (b) of Fig.\ref{fig_1} (the whole 3-dimensional plot have a  radial symmetry in the space domain, and has therefore a bi-conical geometry). A clear X-shaped structure emerges: this non-trivial shape of the spatio-temporal two-photon correlation, that we shall call {\em X-entanglement} can be considered the counterpart, at the level of quantum noise, of the nonlinear X-waves \cite{conti2007}. 
Similar results hold for a variety of phase-matching conditions, corresponding to different crystal orientations, and for a wide range of crystal lengths \cite{caspani2008}. 
\par
\begin{figure}[ht]
%\centering
\includegraphics[width=7.6cm,keepaspectratio=true]{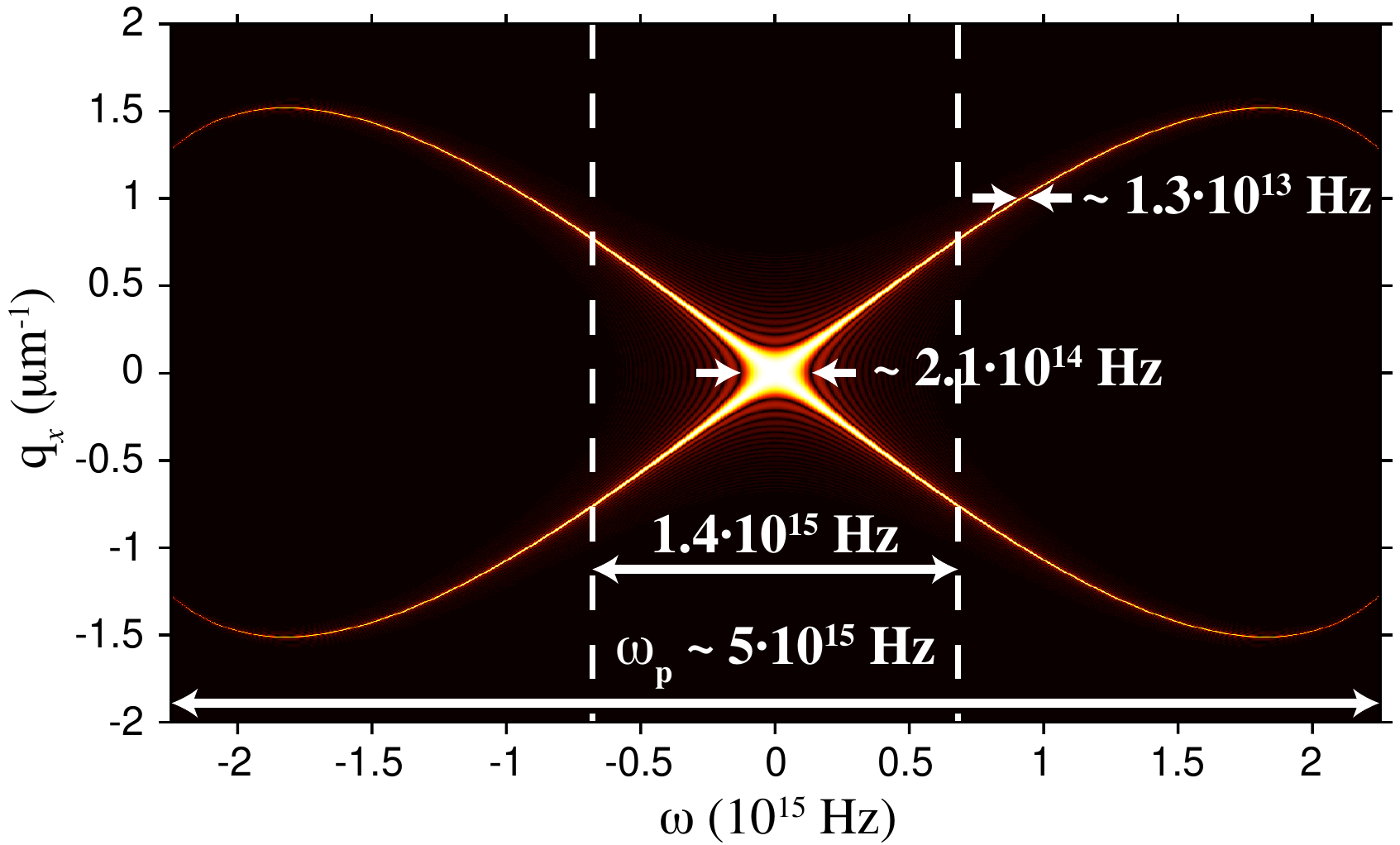}
\caption{(Color online) Plot of $|V(q_x,\omega)|$, showing the phase matching curve. 
The dashed vertical lines show the frequency bandwidth selected by the filter and the arrows indicate the different bandwidths involved (same parameters as in  Fig.\ref{fig_1}). }
% the two-photon coincidences measured at fixed $q$ in the far-field zone have a broad temporal localization determined by the %narrow thickness of the curve (width $\sim 10^13-10^14 $ Hz), while in the near-field zone the temporal localization of two-%photon coincidences is determined in principle by the full bandwidth of phase matching ($\sim 5 \, 10^15 $ Hz), or in practice by %the bandwidth of detection.}
\label{fig_2}
\end{figure}
\begin{figure}[ht]
%\centering
\includegraphics[width=7.6cm,keepaspectratio=true]{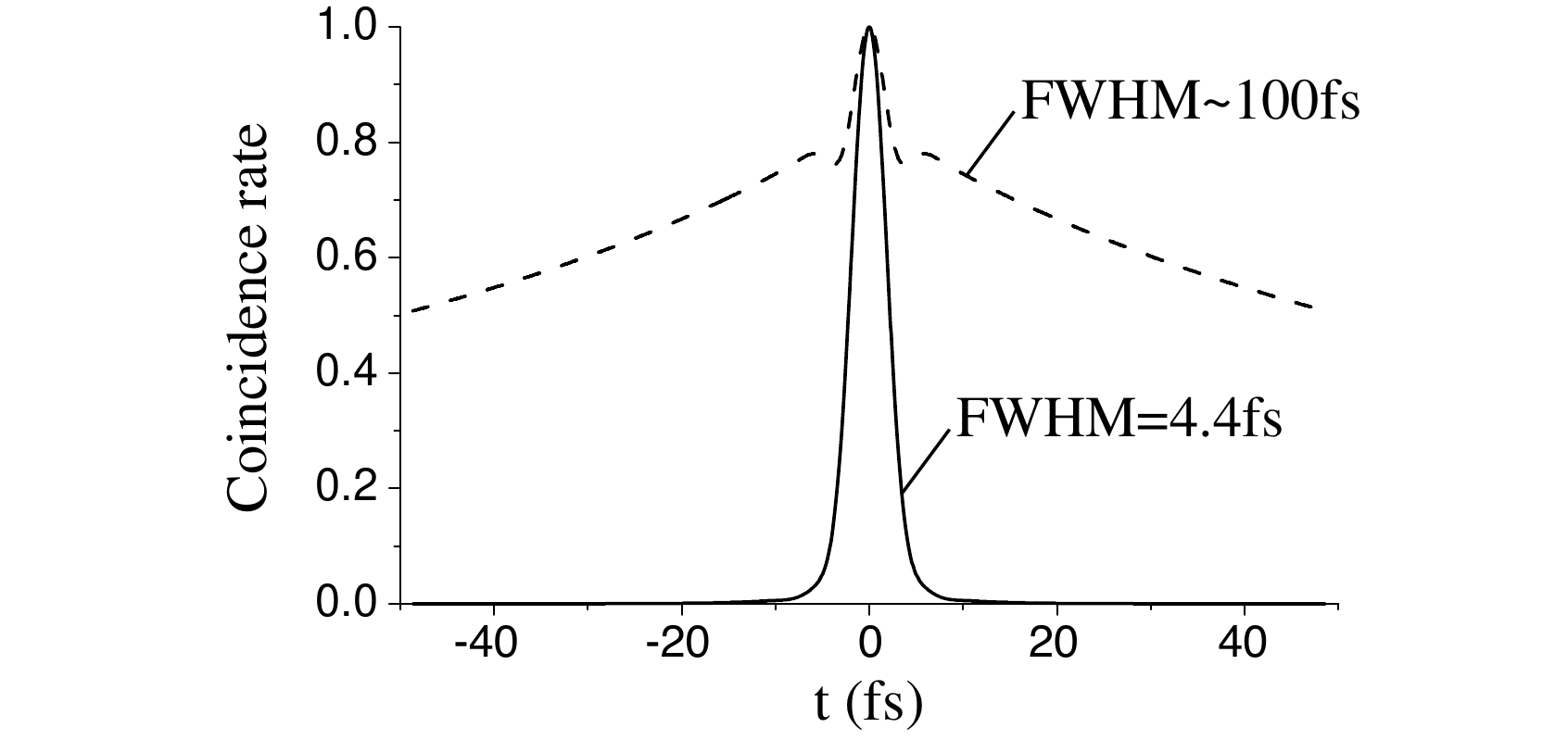}
\caption{Solid line: $ \left| \psipw(\x=0,t)\right|^2$, coincidence rate when photon positions are resolved in the near field. Dashed line: 
$\int  d^2 \x \, \left| \psipw(\x,t)\right|^2$, coincidence rate measured without resolving photon positions}
\label{fig_3}
\end{figure}

A remarkable characteristic of the X-entanglement is the unusually small width of the spatio-temporal correlation peak, which corresponds to a strong relative localization of twin photons both in time and space. The two lower frames of Fig.\ref{fig_1}  plot cuts of the two-photon coincidence rate $|\psipw (\x,t)|^2$ along the temporal and spatial axis, respectively. 
The spatial localization is  remarkable but not impressive, as displayed by  the spatial profile $|\psipw(x,0)|^2$  in  Fig.\ref{fig_1}(d),  which has a FWHM of $\sim 2.9 \,\mu$m.  More impressive, and in a sense, unexpected, is the temporal relative localization of twin photons, which can be appreciated from the temporal profile $ |\psipw(0,t)|^2  $ in  Fig.\ref{fig_1}(c), which is as narrow as  4.4\,fs. Such an ultra-short two-photon localization emerges spontaneously from a nearly monocromatic pump, as a consequence of the ultra-broad bandwidth of PDC phase-matching, which in principle extends over the optical frequency $\omega_p\sim 5\,  10^{15}$ Hz. Notice that,  in order to account for e.g. the finite bandwidth of detection,
in our calculations we include a super-gaussian frequency filter centered at degeneracy. 
%with FWHM=1.4 $10^{15}$ Hz ( wavelength range: $550-950$nm). 
The 4.4 fs width of the temporal peak is in practice determined by the width of this frequency filter, (see Fig.\ref{fig_2}  for a comparative view of the frequency bandwidths involved). \\
%%%%%%%%%%%%%%%%%%%%%%%%%%%%%%%%%%%%%%%%%%%%%%%%%%%%%%%%%%%%%%%%%%%%%%%%%%%%%%%%%%%%%%%%%%%%%%
%%%%%%%%%%%%%%%%%%%%%%%%%%%%%%%%%%%%%%%%%%%%%%%%%%%%%%%%%%%%%%%%%%%%%%%%%%%%%%%%%%%%%%%%%%%%%%%%%%%%%%%%%%%%%%%%%
It is interesting to compare our results with the typical $\sim 100$ fs temporal localization of the coincidence rate measured in the far-field zone, by collecting  twin photons that propagate at symmetric  directions $\q$ and $-\q$, within a small angular bandwidth. In that case, the measured quantity  is  proportional to $| V(\q,t)|^2= |\int \frac{d\omega}{2\pi} \e^{-i\omega t} V(\q,\omega) |^2$. Its temporal width  
is determined by the inverse of the bandwidth of the spectrum $V(\q,\omega)$ at fixed $q$, i.e. the narrow ($10^{13}-10^{14}$Hz) thickness of the curve in Fig.\ref{fig_2}. 
This bandwidth can be roughly evaluated as $\Omega_0$ for $q/q_0<<1$, and as $\Omega_0 q_0/q $ for $q/q_0>1$. Clearly, since $\Omega_0 $ scales as $l_c^{-1/2}$,  the shorter the crystal, the 
stronger the temporal localization in the far-field; however,  a 
far-field localization in the femtosecond range would require a crystal as short as $\sim 50 \, \mu$m, with a strongly reduced down-conversion efficiency. 
Conversely, in our case, the detection of coincidences in the near-field gives in principle access to the full ($\sim 10^{15}$Hz) bandwidth of phase matching even for a long crystal.\\
It is however important to stress  that such an extreme temporal localization of twin photons relies on the ability to resolve their positions in the near-field plane. 
Indeed, a measurement collecting all the photons over the beam cross-section, without discriminating their positions, is characterized by the integrated coincidence rate 
$\int d^2\x \, |\psipw(\x,t)|^2 $, reproduced by the dashed curve in Fig.\ref{fig_3}, which has a width of $\sim100$fs. 
%%%%%%%%%%%%%%%%%%%%%%%%%%%%%%%%%%%%%%%%%%%%%%%%%%%%%%%%%%%%%%%%%%%%%%%%%%%%%%%%%%%%%%%%%%%%%%
%%%%%%%%%%%%%%%%%%%%%%%%%%%%%%%%%%%%%%%%%%%%%%%%%%%%%%%%%%%%%%%%%%%%%%%%%%%%%%%%%%%%%%%%%%%%%%%%%%%%%%%%%%%%%%%%%
This may appear surprising, because in this measurement
%all the photons that propagate at different angles are collected, which,due to the angle-frequency relation imposed by phase %matching, implies that 
all the photons at the different frequencies within the phase-matching are collected. However, the identity $\int d^2\x \, |\psipw(\x,t)|^2=\int d^2 \q |
V (\q,t) |^2 $, shows that in this case the coincidence rate takes the form of an {\em incoherent superposition} of the probabilities of detecting a pair of photons at a given $q$ 
and has therefore the same $\sim 100$fs temporal localization as the far-field coincidence rate at fixed $q$. Conversely, by resolving the near-field positions of twin photons, the measured quantity is  $ |\psipw(0,t)|^2=| 
\int \frac{d^2 \q }{(2\pi)^2}
V(\q,t) |^2 $, which corresponds to a {\em coherent superposition} of the probability amplitudes at a given $q$ (i.e. at a given frequency due to the the angle-frequency relation imposed by phase matching), and therefore allows a stronger temporal localization. 
%%%%%%%%%%%%%%%%%%%%%%%%%%%%%%%%%%%%%%%%%%%%%%%%%%%%%%%%%%%%%%%%%%%%%%%%%%%%
\begin{figure}
%\centering
\includegraphics[width=8cm,keepaspectratio=true]{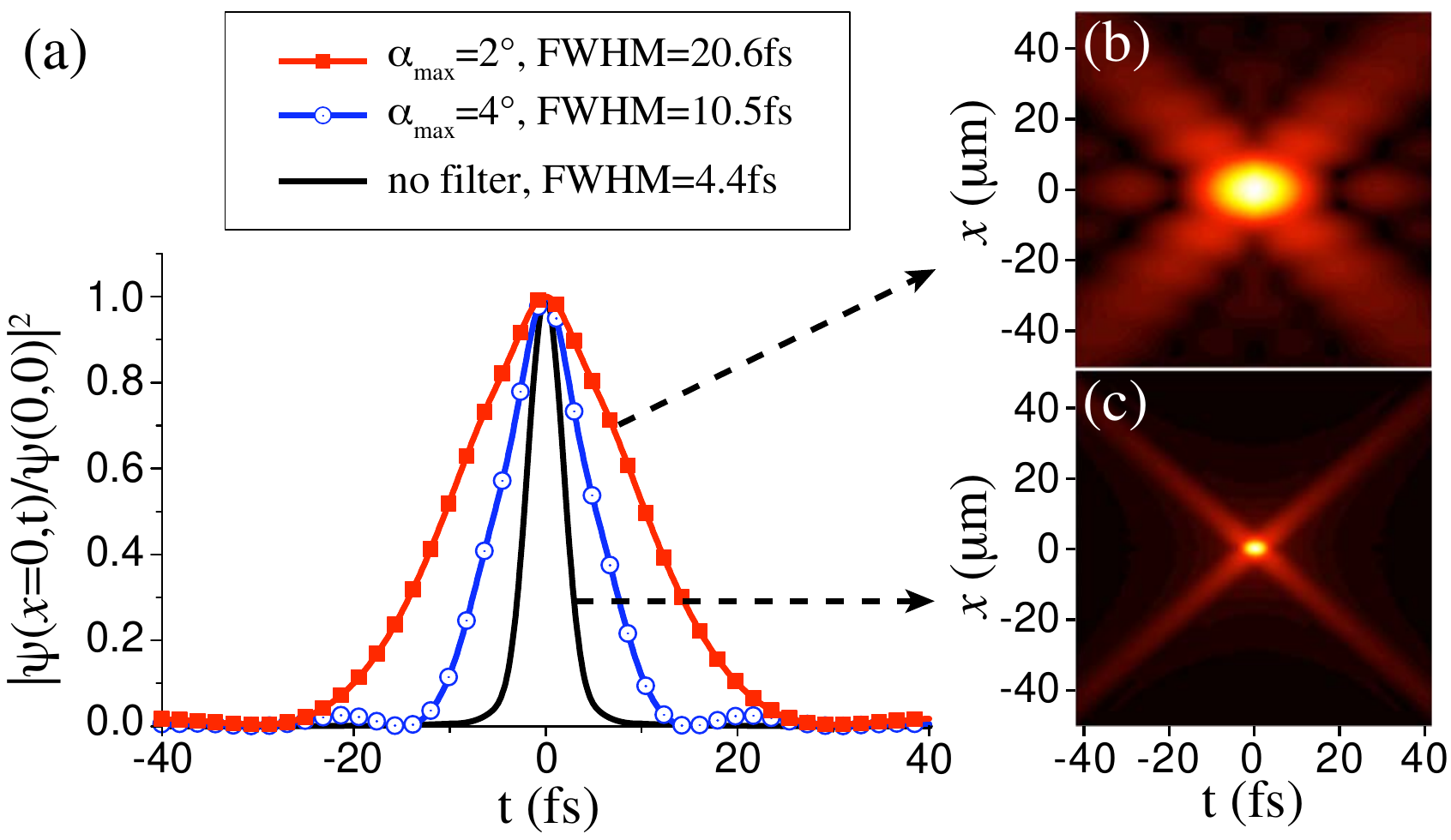}
\caption{(Color online) Effect of spatial filtering on the X-entanglement: (a) Temporal correlation peak $|\psipw(0,t)|^2$ in the presence of a spatial filter, that cuts the angular spectrum at an angle $\alpha_{max}$. The two insets show the full X-correlation for $\alpha_{max}=2^0$ (b), and in the absence of any filter (c).}
\label{fig_filtering}
\end{figure}
%%%%%%%%%%%%%%%%%%%%%%%%%%%%%%%%%%%%%%%%%%%%%%%%%%%%%%%%%%%%%%%%%%%%%%%%%%%%
\par
The non-factorability in space and time of the X-entanglement thus opens the relevant possibility 
of tailoring  the temporal bandwidth of the bi-photons by acting on their spatial degrees of freedom. As a more specific example, let  us consider the effects of spatial filtering on the temporal correlation. 
Let us assume that a  $4f$ lens system is employed to image the near-field of the PDC fluorescence, and that 
a circular aperture of radius $r_a$ is located in the far-field $2f$ plane, acting as a filter that cuts all the angular spectrum at 
$\alpha>\alpha_{max}=\arcsin(r_a/f)$.  Fig. \ref{fig_filtering} shows the effect of such a spatial filter on the temporal correlation peak.   While in the absence of any spatial filter the correlation shows a strong temporal localization, as the angular bandwidth is reduced by spatial filtering, the two-photon correlation broadens in time. This is a clear  effect of the non-factorability of the correlation, because, thanks to the shape of the angular spectrum shown in Fig.\ref{fig_2}, a spatial filter that cuts the angular bandwidth has also the effect of cutting the frequency bandwidth. 
\par
In conclusions,  this work demonstrates  the hyperbolic geometry underlying the two-photon PDC entanglement and its non-factorability with respect to  space and time. As for the X-waves encountered in nonlinear optics, the X-shape of  the bi-photon correlation  is imposed by the phase-matching mechanism governing the PDC process, and following this analogy we coined the name of X-entanglement. The key element of novelty brought by this structure  is its extreme localization, with correlation times and correlation lengths  in the femtosecond and micrometer range, respectively.  The strong temporal localization is determined by the full extent of the PDC bandwidth, rather than by the bandwidth $\sim \Omega_0$ characterizing the PDC far-field. For this reason, a near field measurement scheme able to resolve spatially the coincidences  would provide a powerful tool for high-precision measurements, capable of improving substantially the resolution power in the temporal domain with respect to standard schemes. Furthermore, we have shown how the non-factorability of the structure gives the possibility of tailoring  the temporal entanglement  through manipulations of the spatial degrees of freedom.
\acknowledgments{We thank P.Di Trapani and M. Clerici for useful discussions. We acknowledge the financial support of the FET programme of the EC, under the GA HIDEAS FP7-ICT-221906.}

\end{document}